\documentclass[acmsmall,screen]{acmart}
\usepackage{tikz}
\usepackage{amsmath}
\usepackage{graphicx}
\usepackage[many]{tcolorbox}

\usepackage{amsmath,amssymb,amsfonts}
\usepackage[noend]{algpseudocode}
\usepackage{textcomp}
\usepackage{xcolor,soul}
\usepackage[normalem]{ulem}
\usepackage{amsmath}
\usepackage{mathtools,amssymb,latexsym,amsfonts,stmaryrd}
\usepackage{pbox}
\usepackage{xfrac}
\usepackage{booktabs}
\usepackage{xcolor}
\usepackage{threeparttable}
\usepackage{amsthm}
\usepackage{hyperref}
\usepackage{multirow}
\usepackage{tikz}
\usepackage{mathrsfs}
\usepackage{mathpartir}
\usepackage{algorithm}
\usepackage{url}
\usepackage{syntax}
\usepackage{framed}
\usepackage{flushend}
\usepackage{listings}
\usepackage{moresize}
\usepackage{wrapfig}
\usepackage{xcolor}
\usepackage{colortbl}
 
\usepackage{seqsplit}
\usepackage{alltt}
\usepackage{xspace}
\usepackage{makecell}
\usepackage{subcaption}
\usepackage{diagbox}
\usepackage{rotating}

\usepackage{pifont}

\DeclareMathAlphabet{\mathcal}{OMS}{cmsy}{m}{n}

\usepackage{thmtools}

\declaretheoremstyle[spaceabove=\topsep,notefont=\normalfont\itshape]{mystyle}

\newcommand{\revise}[2]{{\color{red}{\ifx&#1&\else- #1\fi}} {\color{ForestGreen}{\ifx&#2&\else+ #2\fi}}}\renewcommand{\revise}[2]{#2}

\usetikzlibrary{arrows,patterns,decorations.pathreplacing}

\usepackage{microtype}

\newcommand{\F}{Fig.}
\newcommand{\E}{Eq.}
\newcommand{\T}{Table}

\newcommand{\A}{Algorithm}

\newcommand{\ignore}[1]{}

\lstdefinestyle{base}{
  moredelim=**[is][\color{red}]{@}{@},
  escapeinside={<@}{@>}
}

\newcommand{\parh}[1]{\noindent\textbf{#1}}

\usepackage{amssymb}\usepackage{pifont}

\newcommand\DejaVuttfamily{\fontfamily{DejaVuSansMono-TLF}\selectfont }

\lstdefinestyle{base}{
  moredelim=**[is][\color{red}]{@}{@},
  escapeinside={<@}{@>}
}

\lstdefinelanguage
   [x64]{Assembler}     [x86masm]{Assembler} {morekeywords={CDQE,CQO,CMPSQ,CMPXCHG16B,JRCXZ,LODSQ,MOVSXD, POPFQ,PUSHFQ,SCASQ,STOSQ,IRETQ,RDTSCP,SWAPGS, rax,rdx,rcx,rbx,rsi,rdi,rsp,rbp, r8,r8d,r8w,r8b,r9,r9d,r9w,r9b}} 

\usepackage{listings}
\usepackage{fancyvrb}
\usepackage{color}
\definecolor{lightgray}{rgb}{.9,.9,.9}
\definecolor{darkgray}{rgb}{.4,.4,.4}
\definecolor{purple}{rgb}{0.65, 0.12, 0.82}
\definecolor{commentgreen}{RGB}{63,127,95}

\colorlet{myPurple}{blue!40!red}
\definecolor{myOrange}{RGB}{255,192,0}

\lstdefinelanguage{Solidity}{
  keywords={len,delete,int,void,payable, public, event, contract, typeof, new, true, false, catch, function, return, null, catch, switch, var, if, in, while, do, else, case, break,struct,const,socklen_t,sa_familty_t,char,sockaddr},
  keywordstyle=\color{violet}\bfseries,
  ndkeywords={class, export, boolean, throw, implements, import, this},
  ndkeywordstyle=\color{darkgray}\bfseries,
  identifierstyle=\color{black},
  sensitive=false,
  comment=[l]{//},
  escapeinside={(*@}{@*)},          morecomment=[s]{/*}{*/},
  commentstyle=\color{commentgreen}\ttfamily,
  stringstyle=\color{red}\ttfamily,
  morestring=[b]',
  morestring=[b]"
}

\newcommand{\rnum}[1]{\uppercase\expandafter{\romannumeral #1\relax}}

\usepackage{xcolor}

\algnewcommand{\LeftComment}[1]{\Statex \(\triangleright\) #1}

\definecolor{pptbrown}{RGB}{132,60,12}
\definecolor{pptgreen}{RGB}{56,87,35}

\let\OLDthebibliography\thebibliography
\renewcommand\thebibliography[1]{
  \OLDthebibliography{#1}
  \setlength{\parskip}{0pt}
  \setlength{\itemsep}{0pt plus 0.1ex}
}

\definecolor{pptgreen}{RGB}{84,130,53}
\definecolor{pptred}{RGB}{176,35,24}
\definecolor{pptblue}{RGB}{194,214,236}

\definecolor{pptgreen1}{RGB}{78,173,91}
\definecolor{pptred1}{RGB}{192,0,0}

\definecolor{pptyellow1}{RGB}{203,195,167}
\definecolor{pptgreen2}{RGB}{184,192,176}

\newenvironment{packeditemize}{
\begin{list}{$\bullet$}{
\setlength{\labelwidth}{5pt}
\setlength{\itemsep}{0pt}
\setlength{\leftmargin}{\labelwidth}
\addtolength{\leftmargin}{\labelsep}
\setlength{\parindent}{0pt}
\setlength{\listparindent}{\parindent}
\setlength{\parsep}{0pt}
\setlength{\topsep}{0pt}}}{\end{list}}

\newcommand{\tool}{\texttt{PatchAlign}\xspace}
\newcommand{\jtrans}{\texttt{jTrans}\xspace}
\newcommand{\jclap}{\texttt{CLAP}\xspace}
\newcommand{\cebin}{\texttt{CEBin}\xspace}
\newcommand{\insnalign}{\texttt{InsnAlign}\xspace}

\newcommand{\ppd}{PPT\xspace}

\newcommand{\fiber}{\texttt{Fiber}\xspace}
\newcommand{\pdiff}{\texttt{PDiff}\xspace}
\newcommand{\binxray}{\texttt{BinXray}\xspace}

\newcommand{\ps}{\texttt{PS}$^{\text{3}}$\xspace}
\newcommand{\react}{\texttt{React}\xspace}
\newcommand{\robin}{\textsc{Robin}\xspace}
\newcommand{\cidet}{\texttt{CI-Detector}\xspace}

\newcommand{\dsvf}{\texttt{deepseek-v4-flash}\xspace}

\definecolor{lightgreen}{HTML}{b4e5a2}
\definecolor{lightblue}{HTML}{c1e5f5}
\definecolor{lightorange}{HTML}{fbe3d6}
\definecolor{lightyellow}{HTML}{f5f5c1}
\definecolor{lightpurple}{HTML}{f2cfee}
\definecolor{lightred}{HTML}{f5b7b1}
\definecolor{plum}{HTML}{76c2d4}
\definecolor{gray}{HTML}{bfbfbf}
 
\title{Locate, Then Reason: End-to-End Patch Presence Testing in Stripped Binaries}

\author{Huaijin Wang}
\authornote{Corresponding author}
\affiliation{\institution{Shandong University}
\country{China}
}
\email{huaijinwang@sdu.edu.cn}

\author{Tongda Wang}
\affiliation{\institution{Shandong University}
  \country{China}
}
\email{wangtongda@mail.sdu.edu.cn}

\author{Zhibo Liu}
\affiliation{\institution{Nanjing University}
  \country{China}
}
\email{zhiboliu@nju.edu.cn}

\author{Yue Zhang}
\affiliation{\institution{Shandong University}
  \country{China}
}
\email{zyueinfosec@gmail.com}

\author{Shuai Wang}
\affiliation{\institution{Hong Kong University of Science and Technology}
  \country{Hong Kong SAR, China}
}
\email{shuaiw@cse.ust.hk}

\author{Pengfei Hu}
\affiliation{\institution{Shandong University}
  \country{China}
}
\email{phu@sdu.edu.cn}

\author{Jianzhong Liu}
\affiliation{\institution{Shandong University}
  \country{China}
}
\email{liujianzhong@sdu.edu.cn}

\author{Fangtian Zhong}
\affiliation{\institution{Shandong University}
  \country{China}
}
\email{fzhong@sdu.edu.cn}

\author{Shaohua Wang}
\affiliation{\institution{Central University of Finance and Economics}
  \country{China}
}
\email{davidshwang@ieee.org}
 
\begin{document}

\begin{abstract}
Binary patch presence testing (\ppd) determines whether a security fix is
present in a stripped binary. Prior systems often assume that patch-relevant
code is known or successfully matched. Function-level matching is brittle to
compilation changes, inlining, finite model context, and missing usage context;
moreover, similarity suppresses the patch-sized distinction needed for the
verdict. We present \tool, an end-to-end binary
\ppd workflow, which performs \textit{large-scale localization that maximizes recall economically, then the final verdict reasons for patch-specific evidence}.
\tool uses instruction-aligned retrieval and caller-callee context to accurately locate patch-relevant binary code pieces, then screens the shortlist and compares target binary code with vulnerable and patched references using an LLM.
Our experiments indicate that \tool can precisely localize patch-relevant code and determine patch status without a target-location oracle.
\end{abstract}

\maketitle

\section{Introduction}
\label{sec:intro}

Security fixes do not reach every downstream product at the same time~\cite{zhang2021androidpatch,jiang2020pdiff}.
Vendors fork upstream projects and backport selected changes~\cite{zhang2021androidpatch}.
Their products may be available only as stripped binaries, without corresponding source releases or reliable patching information~\cite{pewny2015cross,jiang2020pdiff}.
Binary patch presence
testing (\ppd) asks whether a particular security patch is present in such a
binary and is therefore a basic capability for vulnerability assessment and
patch management~\cite{zhang2018precise,icse2020binxray,jiang2020pdiff,
osprey2021,tosem2023robin,icse2024ps3}.

Binary \ppd is difficult because its signal is both small and unstable. A
patch may change only a few source lines and compile into a handful of
instructions, while the rest of the vulnerable and patched programs remains
nearly identical~\cite{li2017securitypatches,zhang2018precise,icse2020binxray}.
At the same time, compiler choice and optimization can
rewrite those instructions~\cite{junod2015obfuscator}, reshape control flow~\cite{ding2019asm2vec,sem2vec2023}, and move a callee's code
into its callers~\cite{abusabha2026inlining,functioninlining}.
An operational system must consequently solve two coupled
problems: it must locate the code that implements the patch-related behavior
and then decide whether that code implements the vulnerable or patched
semantics~\cite{zhang2018precise,icse2020binxray,marcelli2022machine}.

\begin{tcolorbox}
\textbf{End-to-End Binary \ppd.}
Given a public source patch, source-mapped vulnerable and patched reference
builds, and a stripped target binary, locate the patch-relevant code in the
target---without an oracle for its function or address---and decide whether
the target exhibits the vulnerable or patched behavior.
\end{tcolorbox}

Several binary \ppd systems perform patch verification on an affected target function that is supplied or identified beforehand~\cite{zhang2018precise,icse2024ps3,ase2024react}.
Some take an affected function or entry address directly~\cite{zhang2018precise,icse2024ps3}; others use a
matching front end but cannot test a candidate that was never recovered~\cite{icse2020binxray,tosem2023robin}. This condition is particularly restrictive under
inlining: the affected source function can disappear as a standalone binary
function even though its behavior remains inside a caller~\cite{jiang2020pdiff}.\looseness=-1

Binary code similarity analysis (BCSA) is a natural localization front end since it has been frequently applied to vulnerable function detection~\cite{marcelli2022machine,zuo2019neural,massarelli2019safe,ding2019asm2vec,sem2vec2023,david2014tracelet,wang2026vsim}.
However, conventional function-level similarity does not directly solve
end-to-end patch presence testing.
Consider the tcpdump fix for CVE-2018-14461, which increases two bounds checks from 8 to 12 bytes~\cite{cve201814461}.
In the optimized target binary, the affected function is inlined into a larger caller, so localization must recover its behavior within that caller~\cite{functioninlining}.
Even after finding the correct code, coarse function-level similarity cannot establish whether both corrected checks protect the relevant accesses~\cite{11334478,insnalign2026}.
This example motivates locating the relevant behavior first and then checking the patch-specific conditions.

\smallskip
\parh{Our Approach.}
We present \tool, an end-to-end workflow that separates recall-oriented
localization from patch-conditioned reasoning.
Reference Patch Analysis derives patch-relevant functions and caller-callee contexts from known vulnerable and patched builds, and generates a reusable plan for distinguishing each pair's vulnerable and patched behavior.
Semantic Localization searches CFG windows covering recovered functions and refines candidate rankings using instruction similarity and caller-callee context.
Relevance Screening rejects incidental matches.
Patch-Status Reasoning reuses the corresponding plan with both reference functions and the accepted target code, in assembly or decompiled code.
The final binary verdict aggregates votes from all analyzed functions.
The analysis pipeline requires neither symbols, source mappings, nor a location oracle of the target binary being analyzed.

We evaluate \tool on a rebuilt, balanced version of the \ps benchmark~\cite{icse2024ps3}. The evaluation compares prior systems, separates
abstention coverage from decided-case accuracy, and examines localization,
challenging patches, robustness, component contributions, and cost.

This paper makes the following contributions:
\begin{packeditemize}
\item We identify limitations overlooked when binary \ppd is treated only as
function comparison: prior systems assume patch-relevant code is available,
while function-level matching is brittle under compilation changes such as
inlining; high similarity alone does not establish whether the security fix
is present.

\item We construct \tool, an end-to-end binary \ppd workflow that needs no
target function or address.
It combines recall-oriented, context-aware, and lightweighted semantic localization to locate patch-relevant binary code in the target binary;
then it invokes relatively expensive reasoning model to patch presence reasoning.

\item We perform a comprehensive evaluation for \tool on a balanced benchmark with real-world CVEs against prior systems, covering end-to-end \ppd, localization accuracy, and overhead measurement, demonstrating the essence of end-to-end \ppd for binaries and the effectiveness of our new workflow.
\end{packeditemize}
 \section{Related Work}
\label{sec:related}
\label{sec:background}

\subsection{Existing Binary Patch Presence Testing}
\label{sec:bg:ppd}
\label{sec:related:ppd}

Binary \ppd determines whether a specific security patch is
present in a target binary. Its inputs normally include the source-level
patch, vulnerable and patched reference builds, and a target binary whose
patch status is unknown; its output is a vulnerable or patched verdict.
This task is narrower than general vulnerability search: it must
recognize the small behavioral distinction introduced by one known patch.

Existing systems capture this distinction at several representations.
\fiber derives compact syntactic and semantic signatures from source-level
changes and searches for them in a localized target control flow graph (CFG)~\cite{zhang2018precise}.
\binxray maps vulnerable and patched basic blocks and represents their
difference with basic-block traces~\cite{icse2020binxray}. \pdiff compares
patch-related semantic summaries for downstream kernels~\cite{jiang2020pdiff},
whereas Osprey uses inter-basic-block data-flow slices without symbolic
execution~\cite{osprey2021}. \robin verifies candidates using constrained
symbolic execution and concrete behavior~\cite{tosem2023robin}. More recent
systems seek greater robustness across compilation configurations: \ps uses
symbolic emulation and semantic signature equivalence~\cite{icse2024ps3}, and
\react lifts binaries to an intermediate representation and checks refined
semantic features with an SMT solver~\cite{ase2024react}.\looseness=-1

Despite this progress, three assumptions limit end-to-end use on stripped and
optimized binaries.

\smallskip
\parh{C1: Assumed Patch-Code Location.}
Several systems either accept an affected target function or entry address or
make patch classification conditional on an upstream match. For example,
\fiber's evaluation
uses kernel symbols and leaves similarity-search integration to future work,
while \ps explicitly assumes that the analyzed function's entry address is
known~\cite{zhang2018precise,icse2024ps3}.
\binxray and \robin include lightweight function-matching front ends, but its patch test still operates on the function selected by that stage~\cite{icse2020binxray}.
Thus, not every system omits retrieval, but their patch-classification result generally presupposes that a suitable patch-relevant target function has been found.

\smallskip
\parh{C2: Compilation-Sensitive Evidence.}
Compiler choice and optimization can change instructions, data flow, and CFG
structure even when source semantics are unchanged.
Although \ps, \robin, and \react are designed to improve cross-configuration robustness, their evaluations report residual errors under compiler and optimization changes.
Compilation sensitivity therefore remains a practical limitation rather than
a solved problem~\cite{icse2024ps3,tosem2023robin,ase2024react}.

\smallskip
\parh{C3: Affected Functions Lost to Optimization.}
Function-centric patch tests require a standalone target function on which to
extract or match evidence. Optimization may instead inline the affected
function into one or more callers and remove its entry entirely. In that
case, knowing the source function name is insufficient because the expected
binary object no longer exists.
This case is excluded from the \ps dataset when the affected symbol disappears through inlining, merging, and specialization, and asymmetric inlining is known to invalidate one-to-one binary-function correspondence assumptions~\cite{icse2024ps3,functioninlining,jia2024cross}. \subsection{Binary Code Similarity Analysis}
\label{sec:bg:bcsa}
\label{sec:related:bcsa}

Binary code similarity analysis (BCSA) retrieves code that likely originates
from the same or semantically similar source.
Machine-learning-based (ML-based) methods
encode binary functions as embedding vectors to support efficient queries.
Representative designs include graph neural
networks~\cite{feng2016scalable,xu2017neural}, sequence and random-walk
models~\cite{massarelli2019safe,ding2019asm2vec}, and Transformers with
control-flow-aware encoding~\cite{wang2022jtrans,wang2024clap}.
These methods are valuable as
retrieval mechanisms, but their dominant function-level formulation has four
limitations for \ppd.

\smallskip
\parh{C4: One-to-One Matching under Inlining.}
Conventional BCSA treats recovered function boundaries as semantic boundaries
and generally seeks a one-to-one match. With asymmetric inlining, the
reference may be a standalone callee while the target representation combines
that callee with a much larger caller. The compared embeddings then summarize
different semantic scopes, and there may be no standalone target counterpart
to retrieve~\cite{functioninlining}.
\cidet~\cite{jia2024cross} uses pairwise similarity models~\cite{marcelli2022machine} to classify whether one binary function is cross-inlined into another. Unlike embedding-based approaches~\cite{wang2022jtrans,wang2024clap,insnalign2026}, it neither supports efficient distance-based retrieval nor captures broader semantic-containment relations, making it unsuitable for large-scale semantics containment detection~\cite{wang2026vsim,marcelli2022machine,wang2024cebin}.

\begin{wrapfigure}{r}{0.4\linewidth}
\centering
\includegraphics[width=1.0\linewidth]{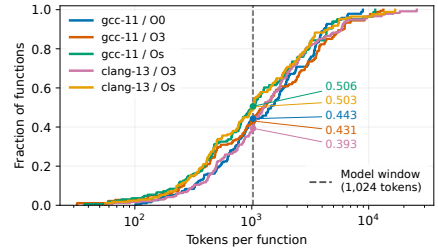}
\caption{Accumulated distribution of patch-relevant function tokens of \ps dataset.}
\label{fig:token-distribution}
\vspace{-5pt}
\end{wrapfigure}

\smallskip
\parh{C5: Finite Representation Context.}
Sequence encoders have a finite input window. Long functions---especially
callers enlarged by inlining---may therefore be truncated before their
patch-relevant instructions are encoded. \insnalign reports retrieval
failures when truncation at a 1,024-token limit discards the
corresponding instructions in optimized functions~\cite{insnalign2026}.
This situation is common, \F~\ref{fig:token-distribution} shows our study in \ps dataset~\cite{icse2024ps3}:
For binaries optimized for sizes (\texttt{gcc-11/-Os} and \texttt{clang-13/-Os}), nearly half of patch-relevant functions exceed the context limit, and this issue is severer for binaries optimized for efficiency.

\smallskip
\parh{C6: Missing Usage Context.}
An isolated function body does not capture how the program uses it. Thin
wrappers, common parsing idioms, and repeated utility code may look alike
locally but call different semantic callees or appear under different callers.
Function-level BCSA is the dominant formulation~\cite{haq2021survey}, and
\insnalign reports that wrapper queries whose callee semantics are invisible
account for most failures in its manual analysis~\cite{insnalign2026}. The
surrounding caller-callee relations can therefore contain discriminating
evidence absent from a single-function representation.

\smallskip
\parh{C7: Patch-Insensitive Similarity.}
BCSA is trained to keep binaries compiled from the same or closely related
source near one another despite compiler variation. That invariance conflicts
with patch presence testing, where a few changed instructions must determine
the verdict. Vulnerable and patched variants consequently remain near twins:
\binxray identifies this as a source of false positives, \robin observes patched
functions being retrieved as vulnerable, and \ps reports that direct
function-level similarity cannot reliably separate the two
variants~\cite{icse2020binxray,tosem2023robin,icse2024ps3}. A high BCSA score
therefore supports code localization, but does not itself establish patch
status.
 \subsection{Instruction Alignment}
\label{sec:bg:insnalign}

\insnalign introduces the instruction-to-source mapping as supervision for binary code
representation learning~\cite{insnalign2026}. Compiler-generated DWARF
information maps binary instructions to their originating source
lines~\cite{dwarf5}. Given two builds of the same source function,
\insnalign treats instructions attributed to the same source line as aligned
positives and adds a multi-positive InfoNCE~\cite{oord2018representation} objective to the usual
function-level contrastive objective. The approach can augment Transformer
backbones such as \jtrans and \jclap, improving function retrieval while exposing
instruction-level similarity as inspectable evidence.
DWARF supplies instruction correspondences during training; the learned
encoder does not require target DWARF at inference.

Formally, let $x=(a_1,\ldots,a_n)$ be an assembly instruction sequence
whose serialized and tokenized input fits within the encoder's context
window. The learned \insnalign encoder maps $x$ to $\mathbf{z}^{x}$, the pooled embedding of the input sequence, and $\mathbf{E}^{x}$, the contextual embeddings of its instructions:
\begin{equation}
(\mathbf{z}^{x},\mathbf{E}^{x})
    = \operatorname{Enc}_{\mathrm{IA}}(x), \qquad
  \mathbf{z}^{x} \in \mathbb{R}^{d_z}, \ \
  \mathbf{E}^{x}=[\mathbf{e}^{x}_1,\ldots,\mathbf{e}^{x}_n]^{\top}
    \in \mathbb{R}^{n\times d_e}.
\label{eq:insnalign-encoder}
\end{equation}

These fine-grained instruction representations support inspecting
patch-relevant correspondences in the target binary. Although they may be
sensitive to some patches, the
instruction-alignment objective alone does not fully address C4-C7.
The study presents patch testing only as a one-CVE feasibility example
and leaves complete binary \ppd to future work.
In contrast, this paper reuses \insnalign's model, while it
focuses on the remaining challenges for real-world, complicated \ppd situations.
  
\begin{wrapfigure}{r}{0.45\linewidth}
\centering
\includegraphics[width=1.0\linewidth]{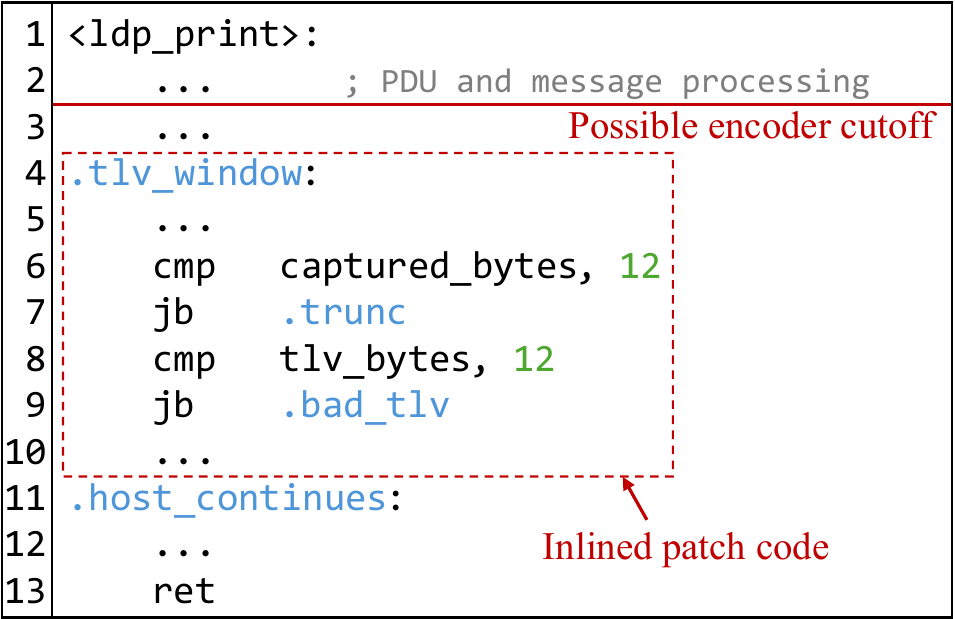}
\caption{Simplified patched assembly for CVE-2018-14461.
    The vulnerable version uses 8 instead of 12 in both bounds checks.
    Symbolic operands and region labels are preserved for readability; they are stripped in the actual analysis.}
\label{fig:motivation-ldp}
\vspace{-5pt}
\end{wrapfigure}

\section{Motivating Example and Design Insights}
\label{sec:challenges}

We revisit the tcpdump example introduced in \S\ref{sec:intro} to derive
four design insights for end-to-end patch presence testing. These insights
address the challenges summarized in \S\ref{sec:related}, from locating
patch-relevant behavior across compilation changes to determining its
patch status.

\smallskip
\parh{Example Details.}
In the \texttt{gcc -O0} reference, \texttt{ldp\_tlv\_print} is a standalone
patch-relevant function. In the \texttt{gcc -O3} target, its logic is inlined
into \texttt{ldp\_print}, whose optimized body also contains PDU and message
processing. The source call chain passes through \texttt{ldp\_pdu\_print};
optimization absorbs these nested calls into the containing function.
\F~\ref{fig:motivation-ldp} illustrates the resulting mismatch between
the reference function and an interior region of the target. Function names
are shown for explanation; target symbols are not required by the analysis.

\smallskip
\parh{Insight 1: Search Within Functions Using Code Windows (C3--C4).}
Inlining preserves the TLV semantics while removing the standalone
\texttt{ldp\_tlv\_print} entry. A patch test that requires this standalone
entry loses its analysis target, and one-to-one BCSA has no standalone
counterpart to match. Comparing the reference callee with all of
\texttt{ldp\_print} also mixes the desired TLV semantics with substantial
caller logic.
The relevant correspondence is between a reference function and code
\emph{within} a target function.
To address this, \textit{retrieval should search code windows
inside recovered functions.} \tool therefore slices CFGs into
windows and searches their embeddings. A window containing relevant TLV
logic can nominate \texttt{ldp\_print} for further analysis.\looseness=-1

\smallskip
\parh{Insight 2: Preserve Complete Coverage with Instruction Embeddings (C5).}
Window-based retrieval must also ensure that the relevant region receives
a representation. The optimized
\texttt{ldp\_print} is a nontrivial parser. With a finite-context encoder such
as \jclap~\cite{wang2024clap}, truncating its serialized body can omit the inlined patch code; those instructions then contribute nothing to its embedding.
Selecting only an entry window would retain this failure.
\textit{Every instruction must participate in representation and matching,
regardless of the containing function's size.} \tool encodes windows covering
the entire function with \insnalign~\cite{insnalign2026}, then scores
candidates using instruction embeddings reconstructed across these windows.

\smallskip
\parh{Insight 3: Enrich Retrieval with Usage Context (C6).}
Complete instruction coverage does not make every local code fragment
distinctive. Length checks and field loads recur across parsers; in the
example, the surrounding LDP PDU and message processing gives the TLV logic
additional meaning. More generally, wrappers and utility functions may be
ambiguous in isolation but distinctive through their callers and callees.
\textit{The usage context of patch-relevant code supplies semantic evidence
that its body alone may lack.} \tool retrieves matches for each reference
patch function and its caller--callee context, then refines candidate rankings
using their semantic and structural consistency while accommodating inlining.

\smallskip
\parh{Insight 4: Reason About Patch-Specific Behavior (C7).}
Even after the correct code is located, the vulnerable and patched TLV
parsers share almost all of their behavior.
High function-level similarity to either reference cannot establish whether both corrected 12-byte checks guard the relevant field accesses on the FT Session path.
Therefore, \textit{patch status requires
reasoning about the behavioral change at the relevant site.}
 \begin{figure*}[t]
    \centering
    \includegraphics[width=1.00\linewidth]{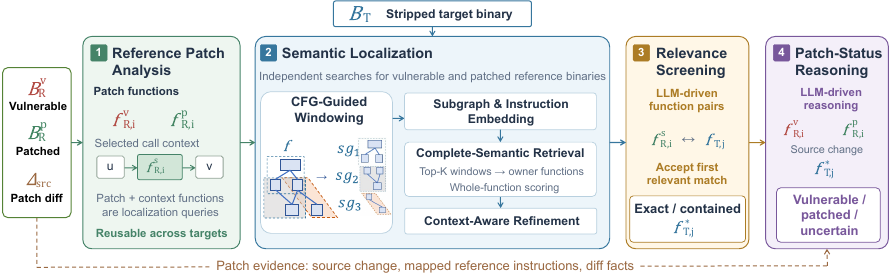}
    \caption{Workflow of \tool.}
    \label{fig:overview}
\end{figure*}

\section{System Design}
\label{sec:design}

\tool first narrows a stripped binary to plausible patch-relevant code, then
uses an LLM for relevance screening and patch reasoning.
As shown in \F~\ref{fig:overview}, \tool takes vulnerable and patched
reference binaries, $B_R^v$ and $B_R^p$, the source change
$\Delta_{\mathrm{src}}$, and a stripped target binary $B_T$. It locates
patch-relevant code and determines its patch status through four stages.

Reference Patch Analysis (\S\ref{sec:design:reference-analysis}) extracts
patch functions, caller-callee contexts, and patch evidence for reuse across
targets. Semantic Localization (\S\ref{sec:design:semantic-search}) uses
instruction similarity and call context to locate candidate target functions,
searching independently from both reference builds. Relevance Screening
(\S\ref{sec:design:relevance}) uses an LLM to reject unrelated candidates.
Finally, Patch-Status Reasoning (\S\ref{sec:design:decider}) uses an LLM to
compare the target code with the source change and both reference
states, returning ``vulnerable'', ``patched'', or ``uncertain''.

\subsection{Reference Patch Analysis}
\label{sec:design:reference-analysis}
\label{sec:design:reference}

Reference Patch Analysis turns the source patch and the two reference builds
into a localization specification, paired vulnerable and patched functions,
one-sided functions, and an analysis plan for distinguishing the paired functions.
Its inputs are the vulnerable
and patched reference-binary representations, $B_R^v$ and $B_R^p$, their
instruction-to-source mappings and function symbols, and the sets of source
lines removed and added by the patch.
Its output is target-independent and can
therefore be reused for every target binary tested against the same reference pair.

\smallskip
\parh{Recovering Reference Patch Functions.}
The analysis first finds functions whose source mappings touch a changed line.
Exact function symbols pair candidates across the two builds.
Because binary \ppd is for detecting known vulnerabilities, particularly for vulnerabilities of reused open-source components~\cite{icse2024ps3,zhang2018precise,ase2024react,duan2019automating}, we can build the vulnerable and patched binaries using identical configuration.
Similar to \ps and \react, we build the reference binaries with \texttt{gcc -O0}.
A paired function is retained only when its normalized assembly differs:
local labels
and hexadecimal immediates are normalized before sequence comparison, and
purely cosmetic differences are discarded. Conservative changes in resolved
internal call targets provide additional patch evidence when call-edge data
is available. A candidate found on only one side is retained rather than being
silently dropped.  We call each retained patch-relevant function in one
reference build a \emph{reference patch function}.  Corresponding reference
patch functions may be paired across the two builds, while unmatched functions
are vulnerable-only or patched-only.  Thus, the analysis does not assume that
a patch-relevant function survives compilation as a one-to-one function in
both references.

\smallskip
\parh{Constructing Context Group.}
For each reference patch function, the analysis
builds a directed call graph from resolved internal calls and terminal tail
transfers.  It then selects a neighborhood centered on that function with
bounded caller and callee depths.  The selector favors nearby nodes with useful
structural evidence, caps high-fanout expansion, suppresses hubs, and enforces
a configurable node budget.
The production setting explores two caller levels and one callee level.
Indirect calls whose destinations cannot be resolved are not invented as edges.

\smallskip
\parh{Generating the Plan for Paired Functions.}
For reference patch functions paired across the two builds based on their function names, an LLM analyzes
the source change and the vulnerable and patched functions to identify the
behavior that distinguishes them.
A second LLM call turns this interpretation into an analysis plan for
recognizing these distinctions in other stripped target code.
The plan specifies the observations supporting each patch-status label
for the paired functions.
The interpretation and plan are generated once per CVE and reference-binary pair
and reused across target binaries in Patch-Status Reasoning
(\S\ref{sec:design:decider}).
In contrast, one-sided reference functions need no plan to distinguish two versions.
When Relevance Screening (\S\ref{sec:design:relevance}) accepts a target match
to a vulnerable-only or patched-only reference function, that function's
presence directly provides a ``vulnerable'' or ``patched'' vote, respectively.

We write $f_{R,i}^s$ for a reference function used as a localization query,
where $s\in\{v,p\}$ identifies the reference build and $i$ identifies the
function. The identifiers cover both reference patch functions and the
context functions; a separate role label distinguishes them.
Some reference patch functions are paired across the two builds, while others are one-sided.
 \subsection{Semantic Localization}
\label{sec:design:semantic-search}
\label{sec:design:identifier}

Semantic Localization performs high-recall search with the encoder from
\insnalign~\cite{insnalign2026}, which produces a pooled representation for coarse retrieval and a contextual representation for every instruction used in fine scoring.
For each reference patch function $f_{R,i}^s$ and the set of functions of stripped target binary $\mathcal{F}_T$, semantic localization retrieves a set of candidate target functions $\mathcal{C}(f_{R,i}^s) \subset \mathcal{F}_T$.

\subsubsection{\textbf{CFG-Guided Windowing}}
\label{sec:cfg-guided-windowing}
\label{sec:design:cfg-guided-windowing}

Binary functions can exceed an encoder's token budget.
Truncation may omit patch code, while linear slicing can separate branches
from the control-flow context needed for high-quality embeddings~\cite{wang2022jtrans,wang2024clap,ding2019asm2vec}.
We therefore slice each function's control-flow graph (CFG) into overlapping
subgraphs that fit the budget.
We denote these windows by $\mathcal{R}(f)$, with $sg_{R,i,k}^s$ and
$sg_{T,j,\ell}$ denoting windows of reference function $f_{R,i}^s$ and
target function $f_{T,j}$, respectively.
Window indices are local to each function and imply no correspondence
across reference builds or target functions.

For a CFG $G=(V,E)$ with entry block $v_0$, let $c(v)$ be the tokenizer's
token count for basic block $v$ and $T$ the encoder's token budget.
Each window selects a block set $W$ satisfying $\sum_{v\in W}c(v)\leq T$
and yields an induced subgraph $H=G[W]$ in $\mathcal{R}(f)$.
\A~\ref{alg:subgraph-slicing} constructs windows using three operations,
distinguishing primary blocks, which advance coverage tracked by $U$,
from context blocks, which allow overlap between slices.

\begin{wrapfigure}{R}{0.54\linewidth}
\vspace{-5pt}
\begin{minipage}{\linewidth}
\begin{algorithm}[H]
{\scriptsize
\caption{CFG-guided windowing}
\label{alg:subgraph-slicing}
\begin{algorithmic}[1]
\Require CFG $G=(V,E)$; entry $v_0$; token costs $c$;
         token budget $T$; predecessor depth $p$
\Ensure Ordered slice collection $\mathcal{S}$

\State $U\gets\emptyset$; $\mathcal{S}\gets[\,]$
    \Comment{primary coverage and output slices}
    \While{$U\neq V$}
        \State $s\gets\Call{ChooseSeed}{V,U,v_0}$
        \State $C, \tau \gets\Call{BackwardContext}{G,s,c,T,p}$
        \label{line:backward}
        \State $P\gets\Call{ForwardPrimary}{G,s,U,C,c,T-\tau}$
        \label{line:forward}
        \State $W\gets\operatorname{set}(P)\cup C$;
               $H\gets G[W]$
        \label{line:subgraph-extraction}
        \State $B\gets\{u\in W:\operatorname{succ}(u)\cap W=\emptyset
          \land u\text{ ends in a jump or branch}\}$
          \label{line:boundary}
        \State $\mathcal{S}.\operatorname{append}
          (H,s,\operatorname{set}(P),C,B)$
        \State $U\gets U\cup\operatorname{set}(P)$
        \Comment{update primary coverage}
    \EndWhile
    \State \Return $\mathcal{S}$

\end{algorithmic}
}
\end{algorithm}
\end{minipage}
\vspace{-5pt}
\end{wrapfigure}

\smallskip
\parh{Choosing a Seed.}
\textsc{ChooseSeed} selects $v_0$ if uncovered; otherwise, it selects the
lowest-address block in $V\setminus U$.
Each seed starts a new primary list, guaranteeing progress.
A block exceeding $T$ is split linearly into slices of at most $T/2$ tokens
until fully covered.

\smallskip
\parh{Collecting Backward Node.}
\textsc{BackwardContext} searches the reversed CFG from the seed to depth $p$
using breadth-first search (BFS), considering predecessors in address order
at each depth.
It adds a predecessor to $C$ if it fits within $T$ alongside the seed and
selected context; otherwise, it skips that block and considers others.
The operation returns $C$ and its token count $\tau$ (Line~\ref{line:backward}).
Context blocks consume budget without updating $U$.

\smallskip
\parh{Growing Primary Coverage.}
\textsc{ForwardPrimary} performs BFS from the seed in stored successor order,
skipping blocks in $U$ or $C$.
It appends visited blocks to the ordered primary list $P$ until the next
block would exceed the remaining budget $T-\tau$ (Line~\ref{line:forward}).
The fixed traversal order makes slicing reproducible for a given serialized CFG.

With the collected blocks $W$,
we extract the subgraph $H$ (Line~\ref{line:subgraph-extraction})
and record conditional-branch blocks with no successor in the slice as
cut boundaries $B$ (Line~\ref{line:boundary}).

Finally, we serialize each slice's blocks in increasing address order and
rebase internal branch targets to local instruction indices, preserving
resolvable control flow and recording cut boundaries.
Encoders that support cut markers can append an unknown pseudo-instruction
at these boundaries; our model embeds only real instructions and produces
one embedding per slice for coarse retrieval.
 \begin{algorithm}
{\scriptsize
\caption{Complete-Coverage Retrieval}
\label{alg:retrieval}
\begin{algorithmic}[1]
\Require A reference query $f_{R,i}^s$, the target functions $\mathcal{F}_T$, the \insnalign encoder $\operatorname{Enc}_{\mathrm{IA}}$ from \E~\eqref{eq:insnalign-encoder}, and the number $K$ of target subgraphs to retrieve per reference subgraph
\Ensure The candidate set $\mathcal{C}(f_{R,i}^s)$ of target functions
\Function{EmbedSubgraph}{$sg$}
  \State $x \gets \operatorname{Serialize}(sg)$
  \State \Return $\operatorname{Enc}_{\mathrm{IA}}(x)$
  \Comment{$(\mathbf{z}^{sg},\mathbf{E}^{sg})$; \E~\eqref{eq:insnalign-encoder}}
\EndFunction

\Function{EmbedFunction}{$f$}
\Comment{$f$ contains $n$ instructions}
  \State $\mathcal{R}(f) \gets$ \Call{SliceCFG}{$f$}
  \State $\mathcal{Z} \gets []$
  \State $\mathcal{H} \gets [\emptyset, \ldots, \emptyset]$
  \Comment{One set for each instruction in $f$}
  \ForAll{$sg_k \in \mathcal{R}(f)$}
    \State $\mathbf{z}_k, \mathbf{E}_k \gets$ \Call{EmbedSubgraph}{$sg_k$}
    \State $\mathcal{Z}.\operatorname{append}(\mathbf{z}_k)$
    \ForAll{$a \in \operatorname{instructions}(sg_k)$}
      \State $\mathcal{H}[a] \gets \mathcal{H}[a] \cup \{\mathbf{E}_k[a]\}$
    \EndFor
  \EndFor
  \ForAll{$a \in \{1, 2, \ldots, n\}$}
    \State $\mathbf{e}_a \gets \frac{1}{|\mathcal{H}[a]|} \sum_{\mathbf{e} \in \mathcal{H}[a]} \mathbf{e}$
    \label{line:retrieval-overlap-average}
    \Comment{Average across overlapping subgraphs}
  \EndFor
  \State \Return $\mathcal{Z}$, $\mathbf{E}=(\mathbf{e}_1,\ldots,\mathbf{e}_n)$
\EndFunction

\Function{CoarseRetrieval}{$f_{R,i}^s$, $\mathcal{F}_T$, $K$}
  \State $\mathcal{Z}_{R,i}^s, \mathbf{E}_{R,i}^s \gets$ \Call{EmbedFunction}{$f_{R,i}^s$}
  \State $m \gets |\mathcal{Z}_{R,i}^s|$
  \State $\mathcal{Z}^{\text{all}} \gets []$
  \Comment{All target subgraph embeddings}
  \State $\mathbf{Owner} \gets []$
  \Comment{Owner function aligned with each target embedding}
  \ForAll{$f_{T,j} \in \mathcal{F}_T$}
    \State $\mathcal{Z}_{T,j}, \mathbf{E}_{T,j} \gets$ \Call{EmbedFunction}{$f_{T,j}$}
    \ForAll{$\mathbf{z} \in \mathcal{Z}_{T,j}$}
      \State $\mathcal{Z}^{\text{all}}.\operatorname{append}(\mathbf{z})$
      \State $\mathbf{Owner}.\operatorname{append}(f_{T,j})$
    \EndFor
  \EndFor

  \State $\mathbf{M} \gets \operatorname{cosine}(\mathcal{Z}_{R,i}^s, \mathcal{Z}^{\text{all}})$
  \Comment{$\mathbf{M}$ has shape $m \times |\mathcal{Z}^{\text{all}}|$}
  \State $\mathcal{C}(f_{R,i}^s) \gets \emptyset$
  \ForAll{$k \in \{1, 2, \ldots, m\}$}
    \State $\mathcal{Q}_k \gets \operatorname{TopKIndices}(\mathbf{M}[k], K)$
    \label{line:retrieval-topk}
    \Comment{Indices of the top-$K$ target subgraphs for reference subgraph}
    \ForAll{$q \in \mathcal{Q}_k$}
      \State $\mathcal{C}(f_{R,i}^s) \gets \mathcal{C}(f_{R,i}^s) \cup \{\mathbf{Owner}[q]\}$
      \label{line:retrieval-candidate-union}
      \Comment{Add the owner function of the retrieved subgraph}
    \EndFor
  \EndFor
  \State \Return $\mathcal{C}(f_{R,i}^s)$
\EndFunction
\end{algorithmic}
}
\end{algorithm}

\subsubsection{\textbf{Complete-Coverage Retrieval}}
\label{sec:design:c1-ranking}
Algorithm~\ref{alg:retrieval} formalizes complete-coverage candidate retrieval.
It first embeds each function from its CFG subgraphs and then uses all resulting
subgraph embeddings for retrieval across the target binary.

While embedding a function $f$, CFG slicing represents it as a set of
subgraphs $\mathcal{R}(f)$. For each subgraph $sg$,
\textsc{EmbedSubgraph} directly reuses the learned \insnalign encoder
defined in \E~\eqref{eq:insnalign-encoder}.
Here, $\operatorname{Serialize}(sg)$ produces the instruction sequence using
the serialization described in \S\ref{sec:cfg-guided-windowing}.
The encoder returns the pooled subgraph embedding $\mathbf{z}^{sg}$ and
the contextual instruction embeddings $\mathbf{E}^{sg}$; we index the
latter by the original instruction identifiers so that
$\mathbf{E}^{sg}[a]=\mathbf{e}^{sg}_a$ for each instruction $a$ in $sg$.
We denote the
embeddings of all subgraphs of $f$ by $\mathcal{Z}(f)$.  Because these
subgraphs jointly cover the entire function, $\mathcal{Z}(f)$ allows retrieval
to consider the function's complete semantics even when the function exceeds
the encoder's input budget.  The subgraphs can overlap, so an instruction may
receive multiple contextual embeddings.  We reconstruct a single contextual
embedding for each instruction in the function by averaging all of its
subgraph-specific embeddings.  If
$\mathcal{R}_a(f)\subseteq\mathcal{R}(f)$ is the set of subgraphs containing
instruction $a$, Line~\ref{line:retrieval-overlap-average} performs this
averaging over all subgraphs in $\mathcal{R}_a(f)$. For a specific function,
we write $\mathcal{Z}_{R,i}^s=\mathcal{Z}(f_{R,i}^s)$ and
$\mathcal{Z}_{T,j}=\mathcal{Z}(f_{T,j})$, with reconstructed instruction
embeddings $\mathbf{E}_{R,i}^s$ and $\mathbf{E}_{T,j}$, respectively.

For the overall retrieval, we apply this embedding procedure to the
reference query $f_{R,i}^s$ and every target function $f_{T,j}\in\mathcal{F}_T$.
For each reference window $sg_{R,i,k}^s$, we retrieve the $K$ target windows
$sg_{T,j,\ell}$ whose embeddings have the highest cosine similarity.
A target function $f_{T,j}$ becomes a candidate if it owns at least one
retrieved window $sg_{T,j,\ell}$.
Lines~\ref{line:retrieval-topk}--\ref{line:retrieval-candidate-union} select
the top-$K$ target subgraphs and insert their owners into
$\mathcal{C}(f_{R,i}^s)$. This set union removes duplicate functions and
allows different reference subgraphs to retrieve candidates independently.

\smallskip
\parh{Whole-Function Scoring.}
After candidate retrieval, we rank only the functions in
$\mathcal{C}(f_{R,i}^s)$ using their whole-function instruction
embeddings. For a candidate target function $f_{T,j}$, let $I_R$ and
$I_T$ be the instruction sets of $f_{R,i}^s$ and $f_{T,j}$, respectively.
Let $S_{ab}=\cos(\mathbf{E}_{R,i}^s[a],\mathbf{E}_{T,j}[b])$ be the cosine
similarity between reference instruction $a\in I_R$ and target instruction
$b\in I_T$. Their symmetric
Mean Alignment Score (MAS)~\cite{insnalign2026} is
{
\begin{equation}
 \operatorname{MAS}(f_{R,i}^s,f_{T,j})=\frac{1}{2}\left(
 \frac{1}{|I_R|}\sum_{a\in I_R}\max_{b\in I_T}S_{ab}
 +\frac{1}{|I_T|}\sum_{b\in I_T}\max_{a\in I_R}S_{ab}\right).
 \label{eq:c1-mas}
\end{equation}}
Candidates are sorted in descending MAS order.  Thus, pooled subgraph cosine
similarity determines candidate membership, while whole-function MAS
determines the final retrieval ranking.
 \subsubsection{\textbf{Context-Aware Refinement}}
\label{sec:design:context-aware-refinement}
\label{sec:design:context-group}

Whole-function similarity leaves two ambiguities: absolute scores are not
directly comparable across reference functions, and optimization can inline a
reference function into another function, removing its standalone counterpart.
Our key insight is that a plausible target location should both have strong
\emph{relative} semantic support and explain the surrounding caller-callee
relations, allowing changes in function boundaries.  We combine per-reference
score normalization with an inlining-aware graph edit search.
For each reference patch function $f_{R,i}^s$ (not the context functions) and its retained candidate $f_{T,j} \in \mathcal{C}(f_{R,i}^s)$, the lowest cost of their context graph edits reflects the relative semantic support and determines the rank of $f_{T,j}$.\looseness=-1

\smallskip
\parh{Relative Semantic Evidence.}
For this subsection, write $f=f_{R,i}^s$ for a reference query and
$t=f_{T,j}$ for a retained target candidate. Let $s(f,t)$ be their MAS score,
and let $s^+(f)$ and $s^-(f)$ be the greatest and least retained
scores.  When these differ, we define
\begin{equation}
 \operatorname{rel}(f,t)=\frac{s(f,t)-s^-(f)}{s^+(f)-s^-(f)},
 \qquad D(f,t)=1-\operatorname{rel}(f,t).
 \label{eq:c2-relative-calibration}
\end{equation}
The best and least similar retained candidates therefore have mismatch costs
zero and one, respectively.  This expresses support within each reference
ranking, not a probability of correctness.

\smallskip
\parh{Inlining-Aware Context Edits.}
\label{sec:directed-context-edit-distance}
\label{sec:design:inlining}
We adapt graph edit distance (GED)~\cite{grapheditdistance2015} to match a small reference call context
inside the target call graph~\cite{sanfeliu1983ged,neuhaus2006fastged}.
Alongside node substitution, edge substitution, and edge deletion, we add an operation, \texttt{merge\_edge\_to\_node}.
For an observed
reference call $u\rightarrow v$, this operation maps both endpoints to one
target $t$: contracting callee $v$ into caller $u$ models inlining.  The
removed node's other call relations remain to be matched, so a contraction
changes function boundaries without discarding context.
Edit costs reflect the evidence supporting each transformation.  Node
substitution uses $D(f,t)$, making strong relative semantic matches cheap.
Callee inlining costs
\begin{equation}
 C_{\mathrm{inl}}(v\Downarrow u,t)=\sigma(v)D(v,t),
 \label{eq:c2-merge-cost}
\end{equation}
where $\sigma(v)$ is the fraction of profiled instructions with semantic
content.  Thus a merge is inexpensive when the target strongly matches the
removed callee or that callee performs little independent computation.

Edge costs measure call connectivity: a direct target call costs zero,
longer directed paths cost more, and absence of a path within the search
horizon incurs edge deletion.  Reference nodes cannot be arbitrarily deleted;
extra target structure is free because the reference context is sought
\emph{inside} the target graph.

\smallskip
\parh{Context-Based Ranking.}
For each reference patch function $f_{R,i}^s$, we reuse the context group
constructed during Reference Patch Analysis
(\S\ref{sec:design:reference-analysis}) and form target candidate locations
from its members' independently retrieved matches and bounded caller-callee
expansion in the target call graph.  This lets context recover a location that
the patch function alone ranks poorly.  For each candidate, we fix the patch
function's mapping and
use bounded beam search to find a low-cost context explanation.  We rank
candidates directly by the lowest mean edit cost found, normalized by the
number of reference nodes and edges, and pass this ranking to Relevance
Screening.  The original top-MAS candidate is evaluated under the same edit
objective as the other candidates.
  \subsection{Relevance Screening}
\label{sec:design:relevance}

Context-aware refinement produces a ranked list of candidate
target functions $f_{T,j}$ for each reference patch function $f_{R,i}^s$.
A high rank, however, does not prove that a candidate implements the same high-level logic. For
example, shared utility code can produce both strong instruction similarity
and a plausible local call pattern.  Relevance Screening therefore asks an LLM
to determine the relationship between each reference patch function and only
its first $K_R$ ranked candidate target functions.\looseness=-1

Each prompt contains exactly one reference-target pair
$(f_{R,i}^s,f_{T,j})$: the complete code (assembly or decompiled pseudocode) of the reference patch function
from its vulnerable or patched reference build, and the complete code of
the candidate target function. It does not expose semantic similarities or
structural edit costs. The candidate's
own symbol and self-references are replaced with an address-derived anonymous
name, and target source mappings are removed; meaningful names of other called
functions may remain as assembly evidence.  The LLM classifies the relation as
``exact'', ``contained'', or ``wrong''.  ``exact''
means the candidate implements the same high-level function.  ``contained''
means the reference patch function's semantics occur within a different target
function, as can result from inlining, function merging, or function
specialization.
``wrong'' means the candidate neither implements the same high-level
function nor contains the reference patch function's semantics.

Candidates are checked in structural rank order. The first candidate classified
as ``exact'' or ``contained'' becomes the accepted target function
$f_{T,j}^{\star}$ for that reference patch function. The star marks acceptance
and does not change the target function's identity. Lower-ranked candidates
are recorded as skipped without another LLM call.
This choice saves token consumption and it is still reliable;
our experiments in \S\ref{sec:eval:relevance} demonstrates that advanced
LLMs can classify irrelevant code with nearly perfect accuracy.
This early-stop policy concentrates generative cost on ambiguous
high-ranked pairs while preserving an explicit ``wrong'' outcome for
incidental retrieval hits.

\smallskip
\parh{Function Status and Handoff.}
For a one-sided reference patch function, an accepted ``exact'' or
``contained'' match establishes that its behavior is present in the target.
The accepted function therefore contributes a ``vulnerable'' vote if the
reference function is vulnerable-only, or a ``patched'' vote if it is
patched-only.
It needs no further comparison in Patch-Status Reasoning.
 \subsection{Patch-Status Reasoning}
\label{sec:design:decider}
\label{sec:design:ppd}

This stage determines the status of accepted target functions associated
with paired reference patch functions, then aggregates function votes into
the target binary's verdict. One-sided functions already receive their status
votes during Relevance Screening (\S\ref{sec:design:relevance}).

\smallskip
\parh{Reasoning for Paired Functions.}
For an accepted target function $f_{T,j}^{\star}$ associated with paired
reference patch functions $f_{R,i}^v$ and $f_{R,i}^p$, the LLM receives:
\begin{packeditemize}
\item the cached patch interpretation and analysis plan produced by
Reference Patch Analysis (\S\ref{sec:design:reference-analysis});
\item the complete code of both the vulnerable reference function $f_{R,i}^v$
and the patched reference function $f_{R,i}^p$; and
\item the complete anonymized code of $f_{T,j}^{\star}$, represented as
assembly or decompiled pseudocode.
\end{packeditemize}
Following the plan, the LLM first identifies the patch-relevant behavior
within the accepted target function, then checks it against the observations
that distinguish the two reference functions. It returns a function-level
label of ``vulnerable'', ``patched'', or ``uncertain'', with a concise
explanation grounded in the target code and both reference functions.
The same plan is reused whenever this reference pair is tested against
another target; it is not regenerated during target reasoning.

\smallskip
\parh{Aggregating Function Votes.}
\tool collects the votes from all accepted patch-relevant functions:
the status votes assigned to one-sided functions during Relevance Screening
and the labels produced by reasoning for paired functions.
Let $V$, $P$, and $U$ denote ``vulnerable'', ``patched'', and ``uncertain'',
respectively, and let $S \subseteq \{V,P,U\}$ be the set of observed vote labels.
The target binary is classified as ``vulnerable'' if $V \in S$ and
$P \notin S$, and as ``patched'' if $P \in S$ and $V \notin S$.
\tool returns ``uncertain'' if both $V$ and $P$ occur in $S$,
or if $S = \{U\}$.
When no functions are accepted ($S = \emptyset$), \tool treats
the binary as ``not affected''. This outcome is recorded as
\textbf{MISSING} and assigned to the negative class alongside
``patched'' outcomes in the evaluation.\looseness=-1
  \section{Evaluation}
\label{sec:evaluation}

We organize the evaluation around the following research questions.
\begin{packeditemize}
\item \textbf{RQ1 (End-to-end effectiveness):} How accurately does \tool
identify vulnerable and patched target binaries compared with \react and \ps?
\item \textbf{RQ2 (Localization):} How effectively do our enhancements contribute to the localization?
\item \textbf{RQ3 (Relevance decision):} How reliably does the LLM classify
candidate relationships as ``exact'', ``contained'', or ``wrong''?
\item \textbf{RQ4 (Efficiency):} What time, model-query, token, and monetary
cost does \tool incur in its offline and online analysis stages?
\end{packeditemize}

\subsection{Experimental Setup}
\label{sec:eval:setup}

\parh{The \ps-Balanced Benchmark.}
We rebuild the benchmark released with \ps~\cite{icse2024ps3} to remove a
confounding factor.
Its CVE-binary observations are distributed unevenly across CVEs.
Consequently, a micro-average can be dominated by CVEs with long release histories or skewed class distributions.
We call our reconstructed benchmark \ps-Balanced. It retains all 62 CVEs from
the four original projects and builds the vulnerable and patched source
revisions under each configuration. We use only \texttt{gcc-11 -O0} binaries
as references; targets are built with \texttt{gcc-11} and \texttt{clang-13}
at \texttt{-O3} and \texttt{-Os}. Thus, each reference-target setting
contains exactly 62 vulnerable and 62 patched targets. As summarized in
\T~\ref{tab:eval:benchmark}, \ps-Balanced contains 124 reference binaries and
496 target binaries. We retain paired stripped and unstripped artifacts,
using the unstripped copies only to construct and audit ground truth.

\begin{table}[t]
\caption{Statistics of \ps-Balanced. Each CVE contributes one vulnerable and
one patched \texttt{gcc-11 -O0} reference, and four vulnerable and four patched
targets built with GCC and Clang at \texttt{-O3} and \texttt{-Os}.}
\label{tab:eval:benchmark}
\vspace{-5pt}
\centering
{
\scriptsize
\begin{tabular}{@{}lrrrrr@{}}
\toprule
\multirow{2}{*}{\textbf{Project}} &
\multirow{2}{*}{\textbf{CVEs}} &
\multirow{2}{*}{\textbf{Ref. Binaries}} &
\multicolumn{2}{c}{\textbf{Target Binaries}} &
\multirow{2}{*}{\textbf{Total Binaries}} \\
\cmidrule(lr){4-5}
& & & \textbf{Vulnerable} & \textbf{Patched} & \\
\midrule
\texttt{FFmpeg}  & 26 &  52 & 104 & 104 & 260 \\
\texttt{OpenSSL} & 23 &  46 &  92 &  92 & 230 \\
\texttt{tcpdump} & 11 &  22 &  44 &  44 & 110 \\
\texttt{libxml2} &  2 &   4 &   8 &   8 &  20 \\
\midrule
\textbf{Total} & \textbf{62} & \textbf{124} & \textbf{248} &
\textbf{248} & \textbf{620} \\
\bottomrule
\end{tabular}
}
\end{table}

\smallskip
\parh{Baselines.}
We compare \tool with the following three baselines:
\begin{packeditemize}
\item \textbf{\ps.} \ps uses symbolic emulation to extract semantic signatures
from the reference binaries and compares them with a specified target
function~\cite{icse2024ps3}. Our adapter additionally recovers an inlined
reference function through its DWARF-identified caller; we therefore label
this strengthened baseline \emph{\ps with inline recovery}.

\item \textbf{\react{} (symbol).} This configuration uses function symbols to
locate the target function, lifts the reference and target functions to an
intermediate representation, and uses refined semantic features and an SMT
solver to determine patch status~\cite{ase2024react}.
This is the official \react{} artifact released with the paper.

\item \textbf{\react{} (Sem-Loc).} This diagnostic configuration replaces
\react's symbol-based target lookup with our Semantic Localization (\S\ref{sec:design:semantic-search}) results while retaining \react's reference processing and patch-presence testing.
We select \react with our localization front end because it is publicly available and presents a better performance than prior works~\cite{ase2024react,icse2024ps3}.
\end{packeditemize}

\smallskip
\parh{Implementation.}
The Semantic Localization (\S\ref{sec:design:semantic-search}) of \tool uses the InsnAlign-enhanced \jclap checkpoint released by \insnalign~\cite{insnalign2026}, a top-$K=50$ target-window retrieval shortlist for each reference window, and a 1,024-token window budget (i.e., the same setting used by \cebin~\cite{wang2024cebin}).
For each reference patch function, we collect its context group with bounded caller and callee edges of length of two, as previous studies~\cite{wang2026vsim,functioninlining} have demonstrated this setting is sufficient for detecting inlined semantics.
The Context-Aware Refinement search
(\S\ref{sec:design:context-aware-refinement}) uses deterministic bounded
beam search to approximate the minimum directed graph edit cost with beam width 32.

We use the \texttt{deepseek-v4-flash-0731} model with temperature 0, \texttt{max} reasoning effort, and 1M context limit for relevance screening and
patch-status reasoning.
Relevance Screening iterates over at most 10 candidates per reference-side
function. Patch-Status Reasoning analyzes accepted ``exact'' or ``contained''
matches to paired reference functions.
 \subsection{RQ1: End-to-End Effectiveness}
\label{sec:eval:overall}

\T~\ref{tab:eval:overall} compares the four settings with a
\texttt{gcc-11 -O0} reference. Each setting contains the same 62 vulnerable
and 62 patched targets.

\smallskip
\parh{Metrics.}
We treat \emph{vulnerable} as the positive class and report $T$rue and $F$alse $P$osiives and $N$egatives (i.e., $TP$, $FP$, $TN$, and $FN$).
Because abstention is operationally meaningful, we also report the number of \textbf{U}ncertain cases.
\T~\ref{tab:eval:overall} presents the \textbf{P}recision, \textbf{R}ecall, \textbf{F1}, and accuracy (\textbf{Acc.}) of all methods; those metrics are commonly used for measuring the performance of binary \ppd solutions~\cite{icse2024ps3,ase2024react,zhang2018precise,jiang2020pdiff}. They are computed by $\textbf{P}=\frac{TP}{TP+FP}$, $\textbf{R}=\frac{TP}{TP+FN}$, $\textbf{F1}=\frac{2\times P \times R}{P+R}$, and $\textbf{Acc}=\frac{TP+TN}{TP+FP+TN+FN}$, respectively.
We also provide non-missing uncertain count, the \textbf{MISSING} count, and selected accuracy (\textbf{Sel. Acc.}) for diagnosis.\looseness=-1

Given the potential catastrophic effect of missing a known vulnerability, similar to how previous solutions, which treats uncertain cases as vulnerable (positive), we adopt a similar policy~\cite{zhang2018precise,icse2024ps3,ase2024react}:
All \textbf{U}ncertain cases are treated as vulnerable, contributing $TP$ and $FP$.
In contrast, \textbf{MISSING} denotes that the solution does not find the patch-relevant code in the stripped binary, indicating the target is considered not being affected by the vulnerability.
This can be caused by missing symbols, semantic localization failures, and relevance screening mistakes.
MISSING cases contribute to negatives (i.e., $TN$ and $FN$), tending to reduce the recall.

\begin{table*}[t]
\caption{Results for \ps dataset.
Vulnerable is positive, and patched is negative.
\textbf{U} is \# non-missing uncertain outcomes.
\textbf{MISSING} denotes \# targets that tool fails to localize the patch-relevant code.
\textbf{Sel. Acc.} is the decision accuracy excluding U and MISSING cases.}
\label{tab:eval:overall}
\vspace{-5pt}
\centering
{
\scriptsize
\setlength{\tabcolsep}{2.3pt}
\begin{tabular}{@{}llrrrrrrrrrrr@{}}
\toprule
\textbf{Setting} & \textbf{Method} & \textbf{U} & {$TP$} & {$FP$} & {$TN$} & {$FN$} & \textbf{MISSING} & \textbf{P} & \textbf{R} & \textbf{F1} & \textbf{Sel. Acc.} & \textbf{Acc.} \\
\midrule
\multirow{5}{*}{GCC -O0$\rightarrow$GCC -O3} & \ps{}+inline & 0 & 42 & 15 & 47 & 20 & 39 & 73.68 & 67.74 & 70.59 & 81.18 & 71.77 \\
 & \react{} (symbols) & 0 & 44 & 12 & 50 & 18 & 37 & 78.57 & 70.97 & 74.58 & 86.21 & 75.81 \\
 & \react{} (Sem-Loc) & 9 & 60 & 27 & 35 & 2 & 3 & 68.97 & \textbf{96.77} & 80.54 & 79.46 & 76.61 \\
 & \tool{} & 8 & 58 & 4 & 58 & 4 & 3 & 93.55 & 93.55 & 93.55 & 96.46 & 93.55 \\
 & \tool{} (decompile) & 3 & 60 & 2 & 60 & 2 & 6 & \textbf{96.77} & \textbf{96.77} & \textbf{96.77} & \textbf{100.00} & \textbf{96.77} \\
\midrule
\multirow{5}{*}{GCC -O0$\rightarrow$Clang -O3} & \ps{}+inline & 0 & 44 & 22 & 40 & 18 & 31 & 66.67 & 70.97 & 68.75 & 74.19 & 67.74 \\
 & \react{} (symbols) & 0 & 43 & 13 & 49 & 19 & 30 & 76.79 & 69.35 & 72.88 & 81.91 & 74.19 \\
 & \react{} (Sem-Loc) & 10 & 54 & 23 & 39 & 8 & 2 & 70.13 & 87.10 & 77.70 & 77.68 & 75.00 \\
 & \tool{} & 5 & 57 & 3 & 59 & 5 & 2 & 95.00 & 91.94 & 93.44 & 94.87 & 93.55 \\
 & \tool{} (decompile) & 2 & 61 & 1 & 61 & 1 & 2 & \textbf{98.39} & \textbf{98.39} & \textbf{98.39} & \textbf{100.00} & \textbf{98.39} \\
\midrule
\multirow{5}{*}{GCC -O0$\rightarrow$GCC -Os} & \ps{}+inline & 0 & 45 & 19 & 43 & 17 & 32 & 70.31 & 72.58 & 71.43 & 78.26 & 70.97 \\
 & \react{} (symbols) & 0 & 44 & 12 & 50 & 18 & 31 & 78.57 & 70.97 & 74.58 & 83.87 & 75.81 \\
 & \react{} (Sem-Loc) & 12 & 58 & 23 & 39 & 4 & 2 & 71.60 & 93.55 & 81.12 & 82.73 & 78.23 \\
 & \tool{} & 5 & 60 & 2 & 60 & 2 & 2 & 96.77 & \textbf{96.77} & \textbf{96.77} & 98.29 & \textbf{96.77} \\
 & \tool{} (decompile) & 1 & 58 & 1 & 61 & 4 & 8 & \textbf{98.31} & 93.55 & 95.87 & \textbf{100.00} & 95.97 \\
\midrule
\multirow{5}{*}{GCC -O0$\rightarrow$Clang -Os} & \ps{}+inline & 0 & 47 & 23 & 39 & 15 & 28 & 67.14 & 75.81 & 71.21 & 75.00 & 69.35 \\
 & \react{} (symbols) & 0 & 46 & 18 & 44 & 16 & 27 & 71.88 & 74.19 & 73.02 & 78.35 & 72.58 \\
 & \react{} (Sem-Loc) & 11 & 57 & 27 & 35 & 5 & 2 & 67.86 & 91.94 & 78.08 & 77.48 & 74.19 \\
 & \tool{} & 9 & 57 & 5 & 57 & 5 & 2 & 91.94 & 91.94 & 91.94 & 95.58 & 91.94 \\
 & \tool{} (decompile) & 2 & 60 & 1 & 61 & 2 & 4 & \textbf{98.36} & \textbf{96.77} & \textbf{97.56} & \textbf{100.00} & \textbf{97.58} \\
\midrule
\multirow{5}{*}{\textbf{Overall}} & \ps{}+inline & 0 & 178 & 79 & 169 & 70 & 130 & 69.26 & 71.77 & 70.50 & 77.05 & 69.96 \\
 & \react{} (symbols) & 0 & 177 & 55 & 193 & 71 & 125 & 76.29 & 71.37 & 73.75 & 82.48 & 74.60 \\
 & \react{} (Sem-Loc) & 42 & 229 & 100 & 148 & 19 & 9 & 69.60 & 92.34 & 79.38 & 79.33 & 76.01 \\
 & \tool{} & 27 & 232 & 14 & 234 & 16 & 9 & 94.31 & 93.55 & 93.93 & 96.30 & 93.95 \\
 & \tool{} (decompile) & 8 & 239 & 5 & 243 & 9 & 20 & \textbf{97.95} & \textbf{96.37} & \textbf{97.15} & \textbf{100.00} & \textbf{97.18} \\
\bottomrule
\end{tabular}
}
\end{table*} 
\smallskip
\parh{Symbol-Based Patch-Relevant Function Identification.}
For \ps and \react (symbol), the MISSING cases are mainly caused by missing symbols in the target binary, which is a common scenario in real-world binaries due to compilation optimizations like function inlining.
Note that we still provide unstripped binaries for them to use symbols for localization, and stripping binaries can further increase MISSING cases in practice. In contrast, \tool does not rely on symbols and can localize patch-relevant code even in stripped binaries, resulting in limited MISSING cases across all settings.

The small MISSING differences between \ps and \react{} (symbols) come from two
artifact-specific cases. For CVE-2019-13312, \ps cannot construct a
reference signature for the ambiguous static name \texttt{encode\_init} and
labels both target states MISSING under all four configurations.
For CVE-2018-16228, \react{} (symbols) failed to lift the LLVM IR for particular patch-relevant functions.

For official \ps and \react, the number of MISSING cases are not negligible. Given 496 target binaries, over 25\% binaries do not contain the expected function symbols.
As a result, the accuracy decreases significantly when treating those MISSING cases as negatives.
Moreover, our experiments of \react (Sem-Loc) demonstrate that the missing symbols is the superfacial outcome, forcing existing patch-status classifier to analyze the patch-relevant code does not close this gap.\looseness=-1

\smallskip
\parh{Semantic-Localization-Assisted \react.}
\react{} (Sem-Loc) uses our Semantic Localization and Relevance Screening results for selecting binary functions in the target binary, but retains \react's reference analysis and classifier.
It reduces MISSING from 125 to 9 and increases recall and F1 compared with \react{} (symbols).
The evaluation of \react also demonstrates that Semantic Localization is effective for patch-relevant code retrieval, but the final end-to-end performance is still limited by the classifier's ability to distinguish patched and vulnerable states.
The Sel. Acc. of \react (Sem-Loc) of all settings also decreases considerably (from 86.21\% to 79.46\% of \texttt{gcc -O0}$\rightarrow$\texttt{gcc -O3}), indicating that aggressive optimization changes can largely affect the classifier's ability to distinguish patched and vulnerable states, even when the patch-relevant code is successfully localized.

\smallskip
\parh{Results of \tool.}
Overall, \tool reaches 93.95\% accuracy and 93.93\% F1, significantly exceeding the best corresponding baseline results.
When we use decompiled pseudocode for relevance screening and decision-making, \tool achieves 97.18\% accuracy and 97.15\% F1, which is the best among all settings.
Compared with the baselines, \tool's main advantage is precision:
it reduces the aggregate false positives (FP) from over hundred for the baselines to 14 and 5, while retaining over 96\% recall.
Together with small MISSING, these results show that \tool provides a substantially better balance between detecting vulnerable binaries and avoiding false alarms across compiler and optimization changes.
 \subsection{RQ2: Patch-Relevant Code Localization}
\label{sec:eval:localization}

We measure localization performance using Recall@$k$, the fraction of queries
for which at least one correct target function appears among the top-$k$
candidates. This metric is widely used in information retrieval and binary
code similarity analysis (BCSA)~\cite{insnalign2026,haq2021survey}.

\smallskip
\parh{Patch-Relevant Ground Truth.}
Conventional BCSA evaluations generally define a correct match by
source-function identity, often recovered from symbols or function names. Our
ground truth instead captures the target function that implements the
patch-relevant behavior. We treat a target function as correct when it is
either an exact counterpart or a different function containing the reference
patch function's semantics. Inlining is one way this containment can arise.
For instance, the patch-relevant function of CVE-2018-14461 is \texttt{ldp\_tlv\_print} in the reference binaries, which is inlined into \texttt{ldp\_print} in the target;
we therefore count \texttt{ldp\_print} as a correct match.
We construct the ground-truth cache from the patch-analysis inventory
and original unstripped ELF/DWARF binaries, following the use of debug
information in prior evaluation~\cite{insnalign2026,functioninlining,jia2024cross}.

\smallskip
\parh{Experimental Setup.}
We evaluate four successive stages: cosine retrieval with linear windows,
CFG-guided windowing (\S\ref{sec:design:cfg-guided-windowing}), whole-function
MAS (\S\ref{sec:design:c1-ranking}), and context-aware refinement
(\S\ref{sec:design:context-aware-refinement}). Following RQ1, we use
\texttt{gcc -O0} references and the four optimized target settings in
\T~\ref{tab:eval:overall}.
We compare vulnerable references with vulnerable targets and patched
references with patched targets.
Across all 62 CVEs, the \texttt{gcc -O0} reference binaries provide 192
patch-relevant functions (95 from vulnerable binaries and 97 from patched
binaries), yielding $768$ ($=192 \times 4$) localization queries across
the four target settings.
The overall results are the pooled performance across all four settings.
Following prior BCSA evaluations~\cite{wang2022jtrans,wang2024clap,wang2026vsim,insnalign2026,ding2019asm2vec,liu2025keenhash},
we assess mean reciprocal rank (MRR) and use the
one-sided paired $t$-tests on per-query reciprocal ranks and report
Holm-adjusted $p$-values ($p_{\mathrm{H}}$) for successive
improvements~\cite{urbano2019statistical,smucker2007comparison}.
Large $t$ and small $p_{\mathrm{H}}$ reflect significant improvements.
MRR uses the first correct candidate's rank in the retrival list, with zero
when no correct candidate is present.

\begin{wrapfigure}{r}{0.47\linewidth}
\vspace{-20pt}
\captionsetup{type=table,position=top}
\caption{Overall patch-relevant code-localization performance over 768 queries. Significance tests compare each stage with its predecessor.}
\label{tab:eval:localization}
\vspace{-5pt}
\centering
\scriptsize
\setlength{\tabcolsep}{2pt}
\begin{tabular}{@{}lrrrr@{}}
\toprule
\textbf{Method} & \textbf{Recall@1/5/10} & \textbf{MRR} & $\boldsymbol{t}_{\mathrm{MRR}}$ & $\boldsymbol{p}_{\mathrm{H}}$ \\
\midrule
Cosine (linear) & 65.23/76.30/79.17 & 0.7007 & -- & -- \\
Cosine (subgraph) & 70.70/78.78/80.99 & 0.7419 & 4.48 & $4.33\!\times\!10^{-6}$ \\
MAS (\S\ref{sec:design:c1-ranking}) & 84.11/89.19/90.23 & 0.8616 & 10.47 & $6.60\!\times\!10^{-24}$ \\
Context (\S\ref{sec:design:context-aware-refinement}) & 87.11/97.27/97.53 & 0.9146 & 4.99 & $3.69\!\times\!10^{-7}$ \\
\midrule
Context (no-rel) & 23.44/65.23/80.21 & 0.4163 & -- & -- \\
Context (no-inl) & 76.17/85.55/87.50 & 0.8052 & -- & -- \\
\bottomrule
\end{tabular}
\vspace{-5pt}
\end{wrapfigure}
 
\smallskip
\parh{Results.}
As shown in \T~\ref{tab:eval:localization}, all three enhancements increase
pooled recall and significantly improve pooled MRR.
The most notable gains come from MAS, which employs the whole-function semantics for BCSA, demonstrating the importance of high semantics coverage for patch-relevant code localization.
The following context-aware refinement considers the call context of each candidate and further improves recall and MRR.
Compared with MAS, Recall@10 increase from 90.23\% to 97.53\%, a high-recall shortlist for Relevance Screening.

Context (no-rel) replaces $\operatorname{rel}(f,t)$ with the absolute MAS
score $s(f,t)$ in \E~\ref{eq:c2-relative-calibration}, reducing Recall@10
to 80.21\%.
High absolute scores can arise from semantics
shared by multiple candidates; normalization rescales their differences
within each reference ranking so that context edit costs reflect relative
semantic support. Context (no-inl) assigns a prohibitive cost of $10^{12}$
to \texttt{merge\_edge\_to\_node}, reducing Recall@10 to 87.50\%.
Without this operation, a reference caller and
callee must match separate target functions, so inlining can cause correct
matches to be penalized as context mismatches despite preserved semantics.\looseness=-1
 \subsection{RQ3: Reliability of Relevance Screening}
\label{sec:eval:relevance}

Relevance screening (\S\ref{sec:design:relevance}) distinguishes ``exact'' and ``contained'' relationships from ``wrong'' candidates that exhibit incidental similarity in the rankings produced by context-aware refinement (\S\ref{sec:design:context-aware-refinement}).
We evaluate the combined retrieval-and-screening outcome and isolate screening
performance on queries with a correct top-10 candidate.

\cidet~\cite{jia2024cross} is a pairwise binary classifier that distinguishes cross-inlining pairs from non-cross-inlining pairs.
We therefore exclude it from RQ2, which evaluates large-scale function localization and requires extreme efficiency~\cite{wang2026vsim,marcelli2022machine,wang2024cebin}.
It is also unsuitable for relevance screening, which must distinguish exact matches, semantic-containment matches (e.g., those arising from inlining~\cite{functioninlining}, function merging~\cite{rocha2019function}, or function specialization~\cite{cooper1993methodology}), and irrelevant pairs,
while \cidet places all non-cross-inlining relationships in a single negative class.

\smallskip
\parh{Metrics.}
Using the ground truth from \S\ref{sec:eval:localization}, Exact requires a
relevant target whose symbol matches the reference after removing compiler-added
suffixes; Contained requires the reference behavior within a target with a
different base symbol. Both measure query success with the correct relation
label; queries with an exact target belong only to Exact.
Overall accepts either positive label for any relevant target, counting each
query once. Irrelevant measures the fraction of evaluated ground-truth
irrelevant pairs correctly labeled ``irrelevant''.
End-to-end uses all eligible queries; conditional rates require a corresponding
ground-truth candidate in the top-10 (either positive relation for Overall
and Irrelevant).\looseness=-1

\begin{table*}[t]
\caption{Relevance screening results, reported as correct/total (percentage).}
\label{tab:eval:relevance}
\vspace{-5pt}
\centering
{
\scriptsize
\setlength{\tabcolsep}{2pt}
\resizebox{\linewidth}{!}{\begin{tabular}{@{}lrrrrrrrr@{}}
\toprule
\multirow{2}{*}{\textbf{Setting}}
& \multicolumn{4}{c}{\textbf{End-to-end}}
& \multicolumn{4}{c}{\textbf{Conditional on valid top-10}} \\
\cmidrule(lr){2-5}\cmidrule(l){6-9}
& \textbf{Exact} & \textbf{Contained} & \textbf{Overall} & \textbf{Irrelevant}
& \textbf{Exact} & \textbf{Contained} & \textbf{Overall} & \textbf{Irrelevant} \\
\midrule
Assembly & 526/546 (96.34\%) & 205/222 (92.34\%) & 737/768 (95.96\%) & 356/369 (96.48\%) & 526/536 (98.13\%) & 205/213 (96.24\%) & 737/749 (98.40\%) & 217/224 (96.88\%) \\
\addlinespace[2pt]
Decompiled C & 523/546 (95.79\%) & 213/222 (95.95\%) & 743/768 (96.74\%) & 380/380 (100.00\%) & 523/536 (97.57\%) & 213/213 (100.00\%) & 743/749 (99.20\%) & 190/190 (100.00\%) \\
\bottomrule
\end{tabular}}
}
\end{table*}
 
\smallskip
\parh{Results.}
\T~\ref{tab:eval:relevance} shows that decompiled C achieves higher Overall
success and is more reliable on Contained and Irrelevant decisions, while
assembly retains a small Exact advantage. The Contained improvement holds
across all compiler/optimization settings: decompiled C preserves every
assembly success and resolves eight
additional queries. Its stronger rejection of irrelevant candidates also
holds on pairs evaluated by both methods.
Since both representations use identical candidate rankings, these
advantages arise during screening.\looseness=-1

Comparing relations exposes different weaknesses. For assembly, Contained
remains harder than Exact even when a true candidate is retrieved.
Decompiled C succeeds on every retrieved Contained query, so
containment is not its main screening limitation in these experiments.
Its Exact shortfall includes false rejections and Exact-to-Contained label
swaps. Both methods also sometimes accept a valid containing host before the
exact counterpart, preserving Overall success while reducing Exact
success.\looseness=-1

 \subsection{RQ4: Efficiency and Cost}
\label{sec:eval:efficiency}

\T~\ref{tab:eval:cost} separates reusable offline analysis from online \ppd.
Local stages report wall time; LLM stages report summed request elapsed time
and recorded input/output tokens, including reasoning tokens.\looseness=-1

\smallskip
\parh{Implementation and Environment.}
We measure 62 CVEs and 496 targets under the four \texttt{gcc-11/-O0}-reference
settings in \T~\ref{tab:eval:benchmark}, starting from disassembled assembly code.
We use Ghidra to produce the pseudocode for our decompilation pipeline.
The experiment was conducted on a 16-core AMD 9950X3D2 server with 256\,GiB RAM and an NVIDIA GeForce RTX 5090 (32\,GiB).
The LLM was \texttt{deepseek-v4-flash-0731} at temperature zero and max reasoning effort.

\begin{table*}[t]
\caption{Offline and online cost of \tool over 62 CVEs and 496 targets.}
\label{tab:eval:cost}
\vspace{-5pt}
\centering
\scriptsize
\setlength{\tabcolsep}{2.8pt}
\begin{tabular}{@{}lrrrrrrr@{}}
\toprule
\textbf{Phase and component} & \multicolumn{4}{c}{\textbf{Time}} &
\textbf{LLM calls} & \textbf{Input tok.} & \textbf{Output tok.} \\
\cmidrule(lr){2-5}
& \textbf{Total (h)} & \textbf{Mean (s)} & \textbf{p50 (s)} &
\textbf{p95 (s)} & \multicolumn{3}{c}{\textbf{Total / mean}} \\
\midrule
\multicolumn{8}{l}{\emph{Offline: once per CVE and reference pair}} \\
\quad LLM CVE interpretation and plan & 1.88 & 109 & 97 & 233 & 124 / 2 & 303{,}918 / 4{,}902 & 707{,}006 / 11{,}403 \\
\midrule
\multicolumn{8}{l}{\emph{Online: once per target binary}} \\
\quad Windowing + Retrieval & 154.46 & 1{,}121 & 521 & 4{,}182 & -- & -- & -- \\
\quad Context-Aware Refinement & 5.19 & 38 & 27 & 117 & -- & -- & -- \\
\midrule[0pt]
\quad Relevance Screening (assembly) & 91.28 & 663 & 322 & 2{,}223 & 3{,}312 / 6.68 & 34.26M / 69{,}078 & 37.52M / 75{,}645 \\
\quad Relevance Screening (decompile) & 81.17 & 589 & 299 & 2{,}055 & 2{,}385 / 4.81 & 36.00M / 72{,}590 & 34.82M / 70{,}198 \\
\midrule[0pt]
\quad Patch-Status Reasoning (assembly) & 27.04 & 196 & 136 & 607 & 757 / 1.53 & 8.03M / 16{,}195 & 12.28M / 24{,}767 \\
\quad Patch-Status Reasoning (decompile) & 19.59 & 142 & 70 & 516 & 762 / 1.54 & 23.40M / 47{,}170 & 7.59M / 15{,}292 \\
\midrule
\quad \textbf{Online total (assembly)} & 277.98 & 2{,}018 & 1{,}425 & 4{,}997 & 4{,}069 / 8.20 & 42.30M / 85{,}273 & 49.80M / 100{,}413 \\
\quad \textbf{Online total (decompile)} & 260.42 & 1{,}890 & 1{,}246 & 4{,}793 & 3{,}147 / 6.34 & 59.40M / 119{,}760 & 42.40M / 85{,}490 \\
\bottomrule
\end{tabular}
\end{table*}

\smallskip
\parh{Cost Distribution.}
Because \tool (decompile) merely requires the pseudocode of several
relevant functions while screening, the cost of decompiling these
functions is negligible and omitted.
For \tool (decompile), Windowing + Retrieval consumes 59.3\% of online time, followed by Relevance Screening (31.2\%),
Patch-Status Reasoning (7.5\%), and Context-Aware Refinement (2.0\%).
Online costs are right-skewed: each stage's p95 is 6.8$\times$ its median.
With the same localization cost, the decompiled total is 1{,}890s per target, assembly-only analysis takes additional 6.77\% time due to increased output tokens.\looseness=-1

\smallskip
\parh{Assembly and Decompiled Pseudocode.}
Pseudocode-based screening uses 28.0\% fewer queries and 11.1\% less summed time,
but similar input and output token consumption as assembly screening.
Its final reasoning uses similar LLM calls, while the number of input tokens is
2.91$\times$ as that of assembly-based reasoning; in contrast, the output
tokens of using pseudocode is merely 61.8\% of assembly-based reasoning.
The significantly different distribution of input and output tokens indicates that decompilation introduces additional tokens for LLM to analysis and alleviates the reasoning efforts.

For prices $p_i=\$0.15$ and $p_o=\$0.6$ per million input and output tokens of \dsvf~\cite{dpskpriceurl}, the recorded offline cost is $0.304p_i+0.707p_o=\$0.47$; \$0.0076 per CVE on average.
Online LLM costs are $42.30p_i+49.80p_o=\$36.23$ for assembly and $59.40p_i+42.40p_o=\$34.35$ for pseudocode, i.e., \$0.073 (assembly) and \$0.069 (pseudocode) per target binary per CVE on average.

\begin{wrapfigure}{r}{0.55\linewidth}
\vspace{-5pt}
\captionsetup{type=table,position=top}
\caption{Patch-Status Reasoning time costs (s).}
\label{tab:eval:reasoning-cost}
\vspace{-5pt}
\centering
{
\scriptsize
\setlength{\tabcolsep}{2pt}
\begin{tabular}{@{}lrrrrr@{}}
\toprule
\textbf{Method} & \textbf{Total} & \textbf{Mean} & \textbf{p50} & \textbf{p95} &
\textbf{Cost/Search} \\
\midrule
\tool{} (assembly) & 97{,}354 & 196.28 & 136.46 & 607.10 & 16.94\% \\
\tool{} (decompile) & 70{,}521 & 142.18 & 70.47 & 515.76 & 12.27\% \\
\ps{}+inline (target+decision) & 959 & 1.93 & 1.25 & 4.88 & 0.17\% \\
\react{} (lift+decision) & 2{,}306 & 4.65 & 2.74 & 15.21 & 0.40\% \\
\bottomrule
\end{tabular}
}
\vspace{-5pt}
\end{wrapfigure}

\smallskip
\parh{Comparison with Baselines.}
\T~\ref{tab:eval:reasoning-cost} compares target-side patch-status reasoning costs after reusable preparation.
The \ps and \react timers excludes the disassembly and offline signature generation.
\react additionally includes the time to lift the assembly of target binaries to IR.
Overall, the patch-status reasoning cost of \ps and \react is nearly negligible compared with the semantic localization cost.
\tool's final reasoning costs more than either baseline's local matching:
its summed time equals 16.94\% of deterministic localization time for assembly
and 12.27\% for pseudocode.
However, given the accuracy rises from 74.80\% (\react Sem-Loc) to 96.17\% (\tool decompile), these costs are justified.

 \subsection{Case Study: Misleading Decompiled Pseudocode}
\label{sec:eval:case-study}

\begingroup
\setlength{\emergencystretch}{1em}
\parh{Correct Localization.}
For CVE-2021-23841, the correct function ranks first after semantic localization on both reference sides.
The affected OpenSSL function, \texttt{X509\_issuer\_and\_serial\_hash},
is located in a patched \texttt{gcc-11 -O3} target using
\texttt{gcc-11 -O0} references.
The fix checks the result of \texttt{X509\_NAME\_oneline} for null before
\texttt{strlen}~\cite{cve202123841}; this guard is visible in
Listing~\ref{lst:case-study-x509} (Line 5).

\begin{lstlisting}[
  float=!t,
  language=C,
  basicstyle=\ttfamily\scriptsize,
  keywordstyle=\ttfamily\bfseries,
  identifierstyle=\ttfamily,
  commentstyle=\ttfamily\color{darkgray},
  stringstyle=\ttfamily,
  numbers=left,
  numberstyle=\tiny\color{darkgray},
  numbersep=6pt,
  xleftmargin=2em,
  xrightmargin=0pt,
  frame=single,
  columns=fixed,
  basewidth=0.5em,
  keepspaces=true,
  showstringspaces=false,
  literate={},
  moredelim={**[is][\color{blue!60!black}\bfseries]{@S@}{@/S@}},
  moredelim={**[is][\color{red!70!black}\bfseries]{@T@}{@/T@}},
  captionpos=b,
  caption={Lines 11--13 show corresponding calls and arguments. Meaningful
    target types and field names come from Ghidra's library type archive and
    survive function-name anonymization. In lines 12--13, Ghidra
    \textbf{incorrectly assigns} the \texttt{\textcolor{red!70!black}{signature}} and
    \texttt{\textcolor{red!70!black}{sig\_alg}} to the memory locations accessed by the
    \textcolor{blue!60!black}{blue source fields}.
    The LLM trusts these wrong labels and rejects the correct match.},
  label={lst:case-study-x509}
]
/* Patched reference source. */                        | /* Decompiled patched target. */
f = X509_NAME_oneline(                                 | __s = FUN_1a0350(
    a->cert_info.issuer,                               |     *(X509_NAME **)&(a->ex_data).dummy,
    NULL, 0);                                          |     (char *)0x0, 0);
if (f == NULL)  /* patch-added guard */                | if (__s != (char *)0x0) {
    goto err;                                          |     ...
...                                                    |     if (iVar1 != 0) {
                                                       |         ...
                                                       |         if (iVar1 != 0) {
                                                       |             ...
if (!EVP_DigestUpdate(ctx,                             |             iVar1 = FUN_11e8b0(ctx,
    (unsigned char *)@S@a->cert_info.serialNumber.data@/S@,   |                 @T@a->signature@/T@,
    (unsigned long)@S@a->cert_info.serialNumber.length@/S@))  |                 @T@(long)*(int *)&a->sig_alg@/T@);
    goto err;                                          |             if (iVar1 != 0) {
...                                                    |                 ...
\end{lstlisting}

\smallskip
\parh{Why Decompiled-C Screening Fails.}
Ghidra applies an incompatible library definition of \texttt{X509}, rendering
the serial-number data and length as \texttt{signature} and \texttt{sig\_alg}.
{The LLM mistakes these inferred names for different digest inputs}
and rejects the correct function on both reference sides.
With no accepted candidate, no paired-function reasoning is invoked:
\tool treats the target binary as \emph{not affected}, resulting in a MISSING of \T~\ref{tab:eval:overall} and producing a false negative on the vulnerable target.\looseness=-1

\smallskip
\parh{Insight.}
Decompiled C improves overall end-to-end results (\T~\ref{tab:eval:overall}),
but incorrect annotations introduce additional noise into LLM analysis.
In the assembly-based run, LLM matches the serial-number accesses at
offsets \texttt{0x8} and \texttt{0x10}, accepts the function as \emph{exact},
and subsequent Patch-Status Reasoning follows the reference analysis plan
and correctly returns \emph{patched}.
Different reference inputs and callee-name handling make this an illustrative
comparison. The case motivates treating inferred types as fallible evidence
and selectively checking disputed accesses against assembly offsets and data
flow.
\par
\endgroup

  \section{Threats to Validity}
\label{sec:threats}

\parh{External Validity.}
The benchmark contains 62 CVEs from four C/C++ projects, evaluated with
GCC \texttt{-O0} references and four optimized target settings.
Generalization to other projects, build configurations, patch distributions,
and LLMs remains unverified. Balancing target labels enables controlled
comparison but does not reproduce deployment prevalence or equalize project
contributions.\looseness=-1

Our evaluation is limited to \texttt{x86\_64} binaries because the current
embedding model supports only this architecture, consistent with the
benchmarks used by \ps and \react~\cite{icse2024ps3,ase2024react}.
Extending \tool to other architectures would require compiling vulnerable
and patched reference binaries for those architectures and training
compatible embedding models; performance in these settings remains
unverified. Moreover, LLM reasoning over assembly from less commonly used
architectures, such as MIPS, may be less reliable, potentially reducing
the accuracy of both Relevance Screening and Patch-Status Reasoning.

\smallskip
\parh{Internal Validity.}
Public repositories, CVEs, and patches may occur in LLM training
data~\cite{deng2024contamination}. Such prior knowledge is compatible with
\tool's 1-day testing setting, which explicitly supplies the source patch;
successive model updates can extend this knowledge to newly disclosed
vulnerabilities. We also compile and strip target binaries locally under
different compiler and optimization settings, making prior exposure to the
exact evaluated assembly less likely. Even with knowledge of the fix, the
LLM must determine whether the target implements it.
 \section{Conclusion}
\label{sec:conclusion}

\tool enables end-to-end patch presence testing in stripped binaries by
combining semantic code localization with LLM-driven reasoning about
patch-specific behavior. Semantic Localization achieves full instruction
coverage within the encoder's limited context size through
control-flow windows. It further considers how functions are used by their
callers and callees, improving retrieval recall when compiler optimizations
reshape function boundaries. Guided by the source patch and both vulnerable
and patched references, LLM-driven reasoning verifies the relevance of
retrieved code and examines the behavioral distinctions that determine patch
status. Together, these capabilities connect robust code retrieval with
precise reasoning about small security fixes, without requiring target
symbols or a known patch location.

\section*{Data Availability}
To support reproducibility, we commit to publicly releasing the data, code, models, and instructions to reproduce the evaluation results upon publication.

\bibliographystyle{ACM-Reference-Format}
\bibliography{bib/similarity,bib/sca,bib/machine-learning,
bib/other,bib/decompiler,bib/symbolic-execution,bib/analysis,bib/sw,bib/ppd}

\end{document}